\documentclass[9pt,conference]{IEEEtran}
\usepackage{dcase2025}

\usepackage{bm} % for bold math symbols (incl. Greek letters) with \bm{}
\usepackage{booktabs}
\usepackage{makecell}
\usepackage{siunitx} % for better number alignment

\usepackage{dcase2025,amsmath,graphicx,url,times,booktabs, tabularx}

\usepackage{multicol}
\usepackage{cite}
\usepackage{tikz}
\usepackage{pgfplots}
\usetikzlibrary{shapes,arrows,positioning}

\title{Cross-Attention with Confidence Weighting for Multi-Channel Audio Alignment}

\name{Ragib Amin Nihal$^{1}$,
      Benjamin Yen$^{1,2}$,
      Takeshi Ashizawa$^{1}$,
      Kazuhiro Nakadai$^{1}$
      }
\address{$^{1}$Systems and Control Engineering, Institute of Science Tokyo, Japan \;
$^{2}$RIKEN BDR, Japan
}

\begin{document}

\maketitle

\begin{abstract}
Multi-channel audio alignment is a key requirement in bioacoustic monitoring, spatial audio systems, and acoustic localization. However, existing methods often struggle to address nonlinear clock drift and lack mechanisms for quantifying uncertainty. Traditional methods like Cross-correlation and Dynamic Time Warping assume simple drift patterns and provide no reliability measures. Meanwhile, recent deep learning models typically treat alignment as a binary classification task, overlooking inter-channel dependencies and uncertainty estimation. We introduce a method that combines cross-attention mechanisms with confidence-weighted scoring to improve multi-channel audio synchronization. We extend BEATs encoders with cross-attention layers to model temporal relationships between channels. We also develop a confidence-weighted scoring function that uses the full prediction distribution instead of binary thresholding. Our method achieved first place in the BioDCASE 2025 Task 1 challenge with 0.30 MSE average across test datasets, compared to 0.58 for the deep learning baseline. On individual datasets, we achieved 0.14 MSE on ARU data (77\% reduction) and 0.45 MSE on zebra finch data (18\% reduction). The framework supports probabilistic temporal alignment, moving beyond point estimates. While validated in a bioacoustic context, the approach is applicable to a broader range of multi-channel audio tasks where alignment confidence is critical. Code available on: \url{https://github.com/Ragib-Amin-Nihal/BEATsCA}
\end{abstract}


\section{Introduction}

Multi-channel audio recording systems serve applications ranging from professional spatial audio production to scientific bioacoustic monitoring using automated recording units (ARUs) \cite{hoffman2025biodcase}. These systems deploy multiple synchronized devices to capture spatial information, enable source separation, and provide measurement redundancy. Maintaining precise temporal alignment between recording channels presents a significant technical challenge.
\\
The primary obstacle is clock drift between independent recording devices. Oscillator variations due to manufacturing tolerances, temperature changes, and component aging cause temporal desynchronization that accumulates over time \cite{brandstein2001microphone}.
% Consumer-grade equipment exhibits relative drift rates of 0.2-16.4 ms/min. Professional devices experience drift from temperature effects ($\pm$20-50 ppm) and aging (1-10 ppm/year) \cite{benesty2008microphone}.
This drift is often nonlinear and unpredictable in field deployments with variable environmental conditions. Applications requiring sub-millisecond accuracy, like bioacoustic localization, need post-processing correction \cite{lollmann2018improved}.
\\
\begin{figure}[t]
\centering
\includegraphics[width=\columnwidth]{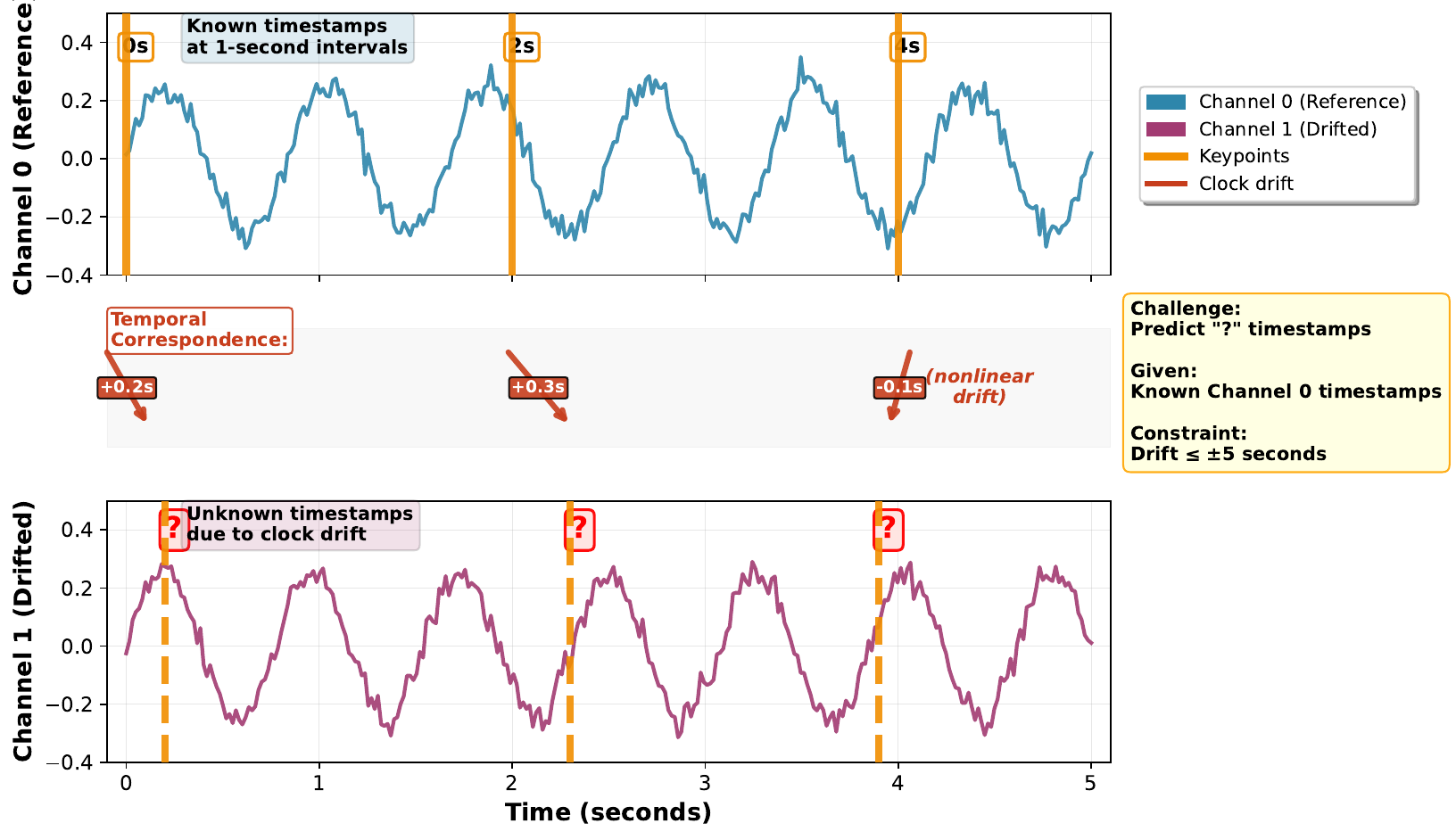}
\caption{Multi-Channel Audio Alignment Problem. Illustration of temporal desynchronization between two audio channels due to nonlinear clock drift. Channel 0 (top) provides reference timestamps at known 1-second intervals, while Channel 1 (bottom) contains corresponding but unknown timestamps (marked ``?"). Arrows indicate temporal correspondences with drift values of +0.2s, +0.3s, and -0.1s, demonstrating the nonlinear nature of clock drift. The challenge is to predict the unknown Channel 1 timestamps given only the Channel 0 reference, with drift constrained to ±5 seconds.}
\label{fig:keypoint_alignment}
\end{figure}
Figure~\ref{fig:keypoint_alignment} shows the alignment task. For temporally drifted stereo recordings, the system must predict corresponding timestamps using sparse keypoint annotations. This requires modeling nonlinear drift while providing confidence estimates.
\\
Traditional methods for multi-channel alignment include cross-correlation techniques. Generalized Cross-Correlation with Phase Transform (GCC-PHAT) \cite{knapp1976generalized} identifies time delays through correlation peaks between channels. These efficient methods assume constant time shifts but fail with nonlinear drift \cite{muller2007information}. Dynamic Time Warping (DTW) handles nonlinear relationships but has $O(N^2)$ complexity and produces unrealistic many-to-one alignments \cite{sakoe1978dynamic}. Modern DTW variants like Memory-Restricted Multiscale DTW reduce computation but lack uncertainty quantification \cite{salvador2007fastdtw}. Speech processing tools such as forced aligners enable automatic phoneme-level segmentation \cite{mcauliffe2017montreal}. However, these approaches primarily address single-modality tasks, leaving multi-channel audio alignment underexplored.
\\
Deep learning approaches have emerged as promising alternatives for temporal alignment. The Audio Spectrogram Transformer (AST) \cite{gong2021ast} uses attention for audio classification, while Conformer architectures \cite{gulati2020conformer} combine convolution and self-attention for temporal modeling. BEATs \cite{chen2023beats} employs iterative pre-training for diverse audio tasks. Self-supervised approaches like wav2vec 2.0 \cite{baevski2020wav2vec} advance speech representation learning. However, existing deep learning alignment systems typically formulate the problem as binary classification—predicting whether audio segments are aligned or misaligned.
\\
This binary classification approach has two limitations that our work addresses. First, methods process channels independently, ignoring correlated clock drift patterns between synchronized devices. Second, binary classification omits reliability estimates needed for scientific applications like bioacoustic analysis \cite{hamada2014fighting}.
\\
Uncertainty quantification has proven valuable in related sequence alignment domains. Methods like GUIDANCE \cite{penn2010guidance} provide confidence scores for alignment regions in bioinformatics applications. Similar probabilistic frameworks are needed for audio alignment, where decisions should be weighted by confidence rather than treated as binary choices.
\\
Current alignment methods cannot model inter-channel dependencies (signal processing) or quantify uncertainty (deep learning). We hypothesize that explicitly modeling inter-channel temporal relationships through cross-attention mechanisms, combined with confidence-weighted scoring that utilizes full prediction distributions, can improve alignment accuracy while providing meaningful uncertainty estimates.
\\
Cross-attention learns relationships between temporal patterns across channels, capturing correlated clock drift. Confidence weighting uses continuous model outputs rather than binary thresholds, enabling certainty-based decisions.
\\
We extend the BEATs encoder \cite{chen2023beats} with cross-attention layers for inter-channel interaction before alignment prediction. Binary counting metrics are replaced by a confidence-weighted function incorporating prediction confidence, top-quartile averaging, and sigmoid-transformed scores. The approach maintains compatibility with existing candidate generation while processing decisions through learned dependencies and probabilistic confidence.
\\
We evaluate our method on the BioDCASE 2025 Task 1 challenge, which provides stereo audio recordings with temporal keypoints from two distinct acoustic domains: field-deployed automated recording units and controlled laboratory recordings. The challenge constrains drift to $\pm$5 seconds and uses mean squared error for evaluation.
\\
Our work contributes:
\begin{enumerate}
    \item Cross-attention mechanisms that model inter-channel temporal dependencies for audio alignment
    \item A confidence-weighted scoring framework using full prediction distributions to quantify alignment uncertainty.
\end{enumerate}
Experimental results show improved performance on the BioDCASE 2025 benchmark.

\section{Problem Formulation}

% We define multi-channel audio alignment as predicting temporal correspondences with uncertainty quantification. Given a stereo audio recording $\mathcal{A} = (\mathcal{A}_0, \mathcal{A}_1) \in \mathbb{R}^{2 \times T}$ that has affected by nonlinear clock drift. The goal is to predict corresponding timestamps between channels while providing confidence scores.
We define multi-channel audio alignment as predicting temporal correspondences with uncertainty quantification. Given a stereo audio recording $\mathcal{A} = (\mathcal{A}_0, \mathcal{A}_1) \in \mathbb{R}^{2 \times T}$ that has been affected by nonlinear clock drift. The goal is to predict corresponding timestamps between channels while providing confidence scores.

\subsection{Mathematical Setup}

Temporal correspondences are established through keypoints $\mathcal{K} = \{k_0, k_1, \ldots, k_{N-1}\}$, where each $k_i = (k_{i,0}, k_{i,1})$ consists of timestamps from Channel 0 (reference) and Channel 1 (drifted). Channel 0 timestamps occur at fixed 1-second intervals: $k_{i,0} = i$ for $i \in \{0, 1, \ldots, N-1\}$.
\\
The clock drift function $\mathcal{D}: \mathbb{R} \rightarrow \mathbb{R}$ maps Channel 0 to Channel 1 timestamps:
\begin{equation*}
k_{i,1} = \mathcal{D}(k_{i,0}) = \mathcal{D}(i),
\end{equation*}
subject to the constraint $|\mathcal{D}(t) - t| \leq 5$ seconds.

\subsection{Training and Inference}

\textbf{Training:} The system accesses complete annotations $\mathcal{K}_{train} = \{(k_{i,0}, k_{i,1})\}_{i=0}^{N-1}$ and learns alignment decisions from audio segments extracted around keypoints. For each keypoint $k_i$, we extract audio segments of duration $\tau = 2$ second:
\begin{align*}
\mathcal{S}_{i,0} &= \mathcal{A}_0[\lfloor k_{i,0} \cdot f_s \rfloor : \lfloor (k_{i,0} + \tau) \cdot f_s \rfloor], \\
\mathcal{S}_{i,1} &= \mathcal{A}_1[\lfloor k_{i,1} \cdot f_s \rfloor : \lfloor (k_{i,1} + \tau) \cdot f_s \rfloor],
\end{align*}
where $f_s$ is the sampling frequency and $\lfloor \cdot \rfloor$ denotes integer indexing.
\\
% \textbf{Inference:} Given only Channel 0 timestamps $\mathcal{K}_{test} = \{k_{i,0}\}_{i=0}^{N-1},$ the system predict Channel 1 timestamps ${\hat{k}{i,1}}_{i=0}^{N-1}$ by evaluating candidate alignments with learned scoring.
\textbf{Inference:} Given only Channel 0 timestamps $\mathcal{K}_{test} = \{k_{i,0}\}_{i=0}^{N-1},$ the system predicts Channel 1 timestamps $\{\hat{k}_{i,1}\}_{i=0}^{N-1}$ by evaluating candidate alignments with learned scoring.

\subsection{Candidate Generation Strategy}

Following the baseline approach, we model the drift as affine: $\mathcal{D}(t) \approx \alpha t + \beta$ where $\alpha$ represents drift rate and $\beta$ represents constant offset. The candidate generation creates a discrete set $\mathcal{C} = \{(\alpha_j, \beta_j)\}_{j=1}^{M}$ by sampling:
\begin{align*}
\alpha_j &\in [1 - \frac{\delta_{max}}{T_{dur}}, 1 + \frac{\delta_{max}}{T_{dur}}] \\
\beta_j &\in [-\delta_{max}, \delta_{max}],
\end{align*}
where $\delta_{max} = 5$ seconds and $T_{dur}$ is audio duration. For each candidate, predicted timestamps are $\hat{k}_{i,1}^{(j)} = \alpha_j \cdot i + \beta_j$, generating candidate sets $\mathcal{K}_j = \{(k_{i,0}, \hat{k}_{i,1}^{(j)})\}_{i=0}^{N-1}$.
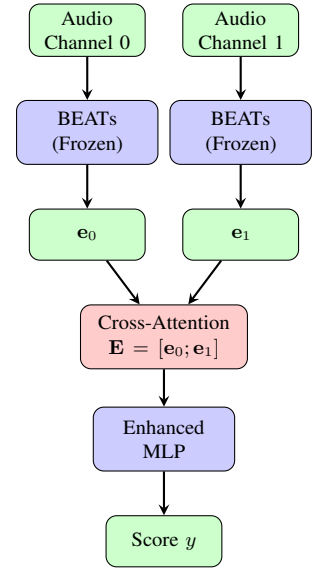
\begin{figure}[t]
\centering
\begin{tikzpicture}[node distance=1.6cm, auto, scale=0.85, transform shape]
    \tikzstyle{block} = [rectangle, draw, fill=blue!20, text width=2cm, text centered, rounded corners, minimum height=1cm]
    \tikzstyle{data} = [rectangle, draw, fill=green!20, text width=1.6cm, text centered, rounded corners, minimum height=0.8cm]
    \tikzstyle{attention} = [rectangle, draw, fill=red!20, text width=2.4cm, text centered, rounded corners, minimum height=1cm]
    \tikzstyle{arrow} = [thick,->,>=stealth]
    
    \node [data] (audio1) {Audio\\Channel 0};
    \node [data, right of=audio1, xshift=0.8cm] (audio2) {Audio\\Channel 1};
    \node [block, below of=audio1] (beats1) {BEATs\\(Frozen)};
    \node [block, below of=audio2] (beats2) {BEATs\\(Frozen)};
    \node [data, below of=beats1] (emb1) {$\mathbf{e}_0$};
    \node [data, below of=beats2] (emb2) {$\mathbf{e}_1$};
    \node [attention, below of=emb1, xshift=1.2cm] (crossatt) {Cross-Attention\\$\mathbf{E} = [\mathbf{e}_0; \mathbf{e}_1]$};
    \node [block, below of=crossatt] (mlp) {Enhanced MLP};
    \node [data, below of=mlp] (output) {Score $y$};
    
    \draw [arrow] (audio1) -- (beats1);
    \draw [arrow] (audio2) -- (beats2);
    \draw [arrow] (beats1) -- (emb1);
    \draw [arrow] (beats2) -- (emb2);
    \draw [arrow] (emb1) -- (crossatt);
    \draw [arrow] (emb2) -- (crossatt);
    \draw [arrow] (crossatt) -- (mlp);
    \draw [arrow] (mlp) -- (output);
\end{tikzpicture}
\caption{System architecture with cross-attention enabling inter-channel interaction before alignment prediction.}
\label{fig:architecture}
\end{figure}
\subsection{Evaluation}

Performance is measured using Mean Squared Error (MSE):
\begin{equation}
\label{eqn}
\mathcal{L}_{MSE} = \frac{1}{N} \sum_{i=0}^{N-1} (k_{i,1} - \hat{k}_{i,1}^*)^2.
\end{equation}
The benchmark combines scores from two datasets: $\mathcal{L}_{final} = \frac{1}{2}(\mathcal{L}_{MSE}(\mathcal{D}_{ARU}) + \mathcal{L}_{MSE}(\mathcal{D}_{zebra}))$.

% Add these packages to your preamble:
% \usepackage{tikz}
% \usetikzlibrary{shapes,arrows,positioning}

\section{Methodology}

We modify BEATs with: (1) cross-attention for inter-channel dependencies, (2) conservative augmentation during training, and (3) confidence-weighted scoring. Figure~\ref{fig:architecture} shows the architecture.

\subsection{Cross-Attention Architecture}

Using BEATs \cite{chen2023beats} with frozen encoders, we combine channel embeddings $\mathbf{e}_0, \mathbf{e}_1 \in \mathbb{R}^{768}$ as $\mathbf{E} = [\mathbf{e}_0; \mathbf{e}_1]$. Multi-head attention \cite{vaswani2017attention} computes,
$
\text{Attention}(\mathbf{Q}, \mathbf{K}, \mathbf{V}) = \text{softmax}\left(\frac{\mathbf{Q}\mathbf{K}^T}{\sqrt{d_k}}\right)\mathbf{V},
$
where $\mathbf{Q}, \mathbf{K}, \mathbf{V} = \mathbf{E}\mathbf{W}_{Q,K,V}$ with learned projections $\mathbf{W}_{Q,K,V} \in \mathbb{R}^{768 \times 768}$.
\\
The enhanced MLP processes attended embeddings $[\mathbf{e}'_0; \mathbf{e}'_1]$ through:
\begin{align*}
\mathbf{h}_1 &= \text{GELU}(\text{LayerNorm}(\mathbf{W}_1[\mathbf{e}'_0; \mathbf{e}'_1] + \mathbf{b}_1)), \\
\mathbf{h}_2 &= \text{GELU}(\text{LayerNorm}(\mathbf{W}_2\mathbf{h}_1 + \mathbf{b}_2)) + \mathbf{h}_1, \\
y &= \mathbf{W}_4(\text{GELU}(\text{LayerNorm}(\mathbf{W}_3\mathbf{h}_2 + \mathbf{b}_3))) + \mathbf{b}_4,
\end{align*}
with layer dimensions 256→128→64→1 and residual connections.

\subsection{Training Procedures}

% \textbf{Data Sampling Strategy:} The baseline randomly samples keypoints during training, potentially causing temporal clustering and poor coverage. We modify this to sample every 20th keypoint per file, ensuring uniform temporal distribution while maintaining computational efficiency. This systematic sampling provides approximately 5\% of available keypoints per file, balancing training diversity with processing constraints.
\textbf{Keypoint Sampling:} To avoid temporal clustering in random sampling, we select every 20th keypoint per file. This covers 5\% of keypoints uniformly while maintaining efficiency.
\\
\textbf{Conservative Data Augmentation:} Applied to 30\% of training samples with channel synchronization:
\begin{itemize}
\item \textit{Amplitude scaling}: Random factor $\alpha \sim \mathcal{U}(0.9, 1.1)$ applied identically to both channels
\item \textit{Gaussian noise}: SNR $\in$ [40, 50] dB
\item \textit{Consistency constraint}: Identical random seed ensures synchronized augmentation across channels
\end{itemize}
The augmentation preserves temporal alignment by applying identical transformations:
\begin{align*}
\tilde{\mathcal{S}}_{i,0} &= \alpha \cdot \mathcal{S}_{i,0} + \mathcal{N}(0, \sigma^2), \\
\tilde{\mathcal{S}}_{i,1} &= \alpha \cdot \mathcal{S}_{i,1} + \mathcal{N}(0, \sigma^2),
\end{align*}
where $\sigma^2$ depends on target SNR. This preserves temporal alignment while improving robustness.

% \textbf{Optimization Strategy:} We replace Adam with AdamW \cite{loshchilov2019decoupled}, adding explicit weight decay regularization that decouples from gradient-based updates:
% \begin{equation}
% \theta_{t+1} = \theta_t - \eta(\hat{\mathbf{m}}_t / (\sqrt{\hat{\mathbf{v}}_t} + \epsilon) + \lambda\theta_t)
% \end{equation}
% where $\lambda = 0.01$ prevents overfitting in the MLP layers. Learning rate scheduling uses ReduceLROnPlateau (factor 0.7, patience 3) to adapt to training dynamics.
\begin{figure*}[t]
\centering
\includegraphics[width=\textwidth]{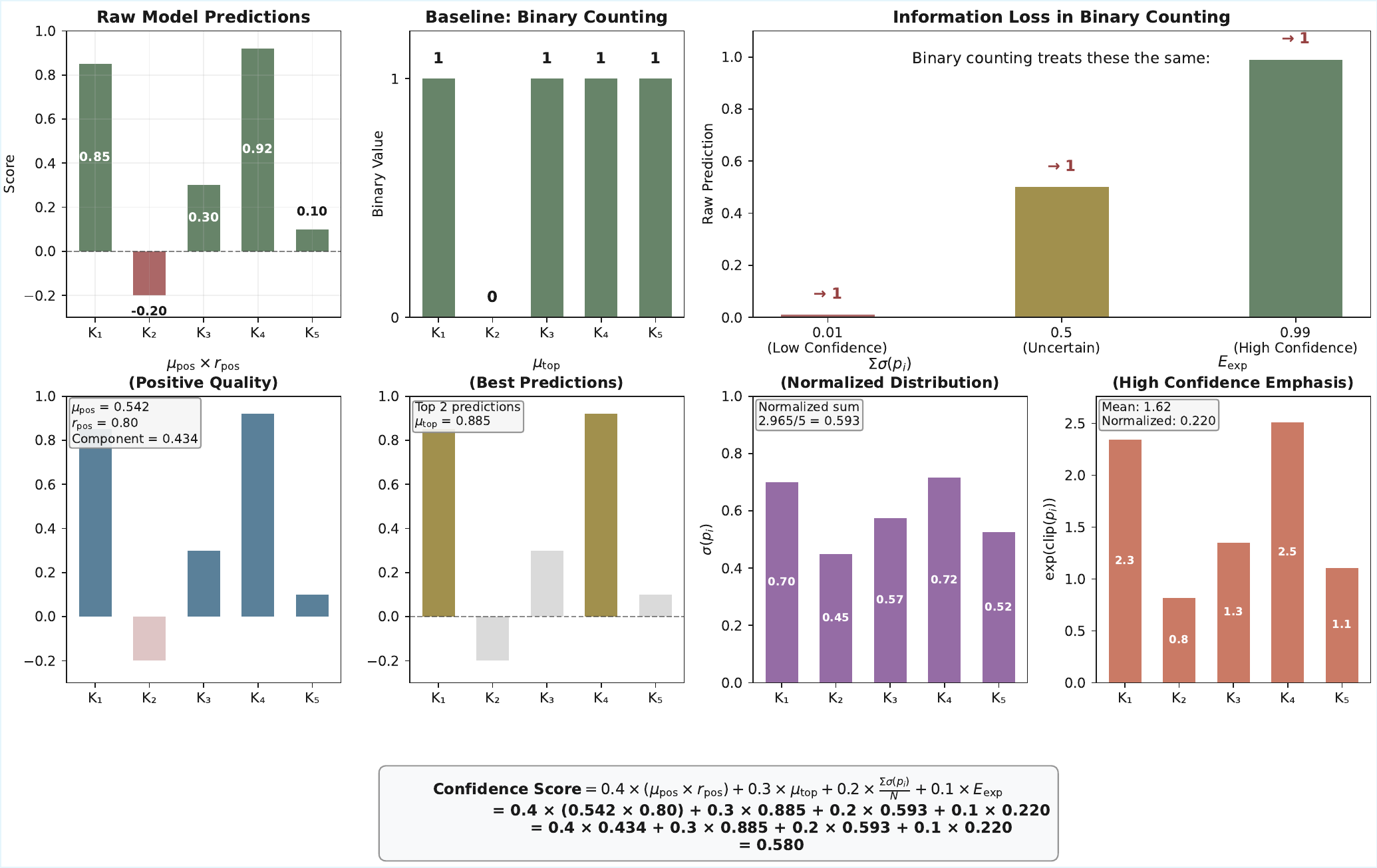}
\caption{Confidence-weighted scoring methodology for multi-channel audio alignment. \textbf{Top row:} Raw model predictions (left) are converted to binary values (0.80 average) in the baseline approach (center), demonstrating information loss where predictions with vastly different confidence levels are treated identically (right). \textbf{Middle row:} Our confidence-weighted approach incorporates four complementary components: positive prediction quality ($\mu_{\mathrm{pos}} \times r_{\mathrm{pos}} = 0.434$), top quartile averaging ($\mu_{\mathrm{top}} = 0.885$), normalized sigmoid distribution ($\Sigma\sigma(p_i)/N = 0.593$), and exponential emphasis of high-confidence predictions ($E_{\mathrm{exp}} = 0.220$). \textbf{Bottom:} The weighted combination (0.4, 0.3, 0.2, 0.1 weights respectively) produces a confidence score of 0.580, demonstrating how confidence weighting utilizes the full prediction distribution compared to binary counting for alignment candidate selection.}
\label{fig:scoring}
\end{figure*}

\subsection{Confidence-Weighted Scoring}

Traditional methods use binary counting, simply summing the number of aligned predictions ($\sum_{i} \mathbf{1}[f_\theta(\mathcal{S}_{i,0}, \mathcal{S}_{i,1}) > 0]$) discarding valuable confidence information. A prediction of 0.01 is treated identically to 0.99, despite vastly different certainty levels. We replace binary counting with a scoring function that incorporates prediction distributions:
\begin{equation}
\label{eqn:score_function}
\mathcal{S}_{conf}(\mathcal{K}_j) = 0.4 \mu_{pos} r_{pos} + 0.3 \mu_{top} + 0.2 \sum_{i} \sigma(p_i^{(j)}) + 0.1 \mathcal{E}_{exp}
\end{equation}
The weights (0.4, 0.3, 0.2, 0.1) were chosen based on intuitive design principles: highest weight for positive prediction quality ($\mu_{pos} \cdot r_{pos}$) as it directly measures alignment confidence, second-highest for top quartile focus ($\mu_{top}$) to emphasize reliable predictions, and lower weights for the supporting probabilistic and exponential components. 
% While this heuristic approach proves effective empirically, principled optimization of these weights represents an important direction for future work.
\\
The components:
\begin{itemize}
    \item \textit{Positive Confidence Weighting} ($\mu_{pos} \cdot r_{pos}$): Average confidence of positive predictions weighted by their prevalence
    \item \textit{Top Quartile Focus} ($\mu_{top}$): Mean confidence of highest-scoring 25\% predictions
    \item \textit{Probabilistic Coverage} ($\sum_{i} \sigma(p_i^{(j)})$): Sum of sigmoid-transformed probabilities
    \item \textit{Exponential Amplification} ($\mathcal{E}_{exp}$): $e^{4(p-0.5)}$ for $p>0.5$, else 0
\end{itemize}
% \textbf{Positive Confidence Weighting} ($\mu_{pos} \cdot r_{pos}$): Measures both the average confidence of positive predictions and their prevalence, emphasizing candidates where the model is both confident and consistent in predicting alignment.
% \\
% \textbf{Top Quartile Focus} ($\mu_{top}$): Concentrates on the highest-confidence predictions, as these likely represent the most reliable temporal correspondences regardless of overall distribution.
% \\
% \textbf{Probabilistic Coverage} ($\sum_{i} \sigma(p_i^{(j)})$): Applies sigmoid transformation to convert raw scores into probabilities, providing a normalized measure of overall alignment likelihood across all keypoints.
% \\
% \textbf{Exponential Confidence Amplification} ($\mathcal{E}_{exp}$): Exponentially weights high-confidence predictions while suppressing low-confidence ones, creating a sharp distinction between reliable and unreliable alignments.
Figure~\ref{fig:scoring} illustrates the scoring approach with an example.

\subsection{Inference Pipeline}

\textbf{Candidate Generation:} Following the baseline, we assume affine drift $\mathcal{D}(t) \approx \alpha t + \beta$ and sample parameters:
$\alpha_j \in [1 \pm 5/T_{dur}]$, $\beta_j \in [-5, 5]$, generating candidates $\hat{k}_{i,1}^{(j)} = \alpha_j \cdot i + \beta_j$.

\textbf{Selection:} For each candidate $\mathcal{K}_j$: (1) extract 2-second segments at predicted locations, (2) compute cross-attention scores, (3) calculate $\mathcal{S}_{conf}(\mathcal{K}_j)$, (4) select maximum-scoring candidate.
% This methodology addresses key limitations: independent channel processing (via cross-attention) and binary decision-making (via confidence weighting), while maintaining computational efficiency through frozen encoders and efficient scoring operations.
\begin{table}[t]
\centering
\caption{BioDCASE 2025 Challenge Results (Test Set)}
\label{tab:competition_results}
\begin{tabular}{l S S S}
\toprule
\textbf{Method} & \textbf{ARU} & \textbf{Zebra Finch} & \textbf{Average} \\
\midrule
Nosync               & 0.85  & 1.84  & 1.35 \\
Crosscorr            & 1.10  & 5.55  & 3.32 \\
DL Baseline          & 0.62  & 0.55  & 0.58 \\
\textbf{Ours (BEATsCA)} & \textbf{0.14} & \textbf{0.45} & \textbf{0.30} \\
\midrule
\textbf{Improvement vs DL} & \textbf{77.4\%} & \textbf{18.2\%} & \textbf{48.3\%} \\
\bottomrule
\end{tabular}
\end{table}

\begin{table}[t]
\centering
\caption{Validation Set Performance Comparison}
\label{tab:validation_results}
\begin{tabular}{l S S S}
\toprule
\textbf{Method} & \textbf{ARU} & \textbf{Zebra Finch} & \textbf{Average} \\
\midrule
Nosync       & 0.976 & 1.315 & 1.146 \\
Crosscorr    & 6.861 & 10.029 & 8.445 \\
DL Baseline  & 0.516 & 1.262 & 0.889 \\
\textbf{Ours} & \textbf{0.099} & \textbf{0.521} & \textbf{0.310} \\
\midrule
\textbf{Improvement vs DL} & \textbf{80.8\%} & \textbf{58.7\%} & \textbf{65.1\%} \\
\bottomrule
\end{tabular}
\end{table}

\section{Experimental Setup}

\subsection{Datasets and Evaluation Protocol}

% We evaluate our method on the BioDCASE 2025 Task 1 challenge datasets~\cite{hoffman2025biodcase}, which provide stereo audio files with temporal keypoints for two distinct domains. The ARU dataset contains 36 training files and 12 validation files from passive automated recording units deployed in field conditions, exhibiting complex drift patterns due to environmental variations. The zebra finch dataset contains 108 training files and 16 validation files from controlled laboratory recordings of zebra finch vocalizations, showing more consistent but still nonlinear drift patterns.
We use BioDCASE 2025 Task 1 datasets \cite{hoffman2025biodcase} containing stereo audio with temporal keypoints in two domains:
\begin{itemize}
    \item \textbf{ARU dataset:} 36 training files, 12 validation files from field recordings
    \item \textbf{Zebra finch dataset:} 108 training files, 16 validation files from laboratory recordings
\end{itemize}
Both datasets contain keypoints at 1-second intervals with drift constrained to $\pm$5 seconds. The challenge evaluates using MSE (Equation~\ref{eqn}) between predicted and ground-truth Channel 1 timestamps, with final scoring computed as the average MSE across both test datasets. All training uses only the provided data without external resources, ensuring fair comparison with baseline methods.

\subsection{Implementation Details}

\textbf{Model Architecture:} We use the pre-trained BEATs encoder (BEATs\_iter3\_plus\_AS2M\_finetuned\_on\_AS2M\_cpt2.pt) with frozen parameters during training. The cross-attention module employs 8 attention heads with 768-dimensional embeddings. The enhanced MLP processes 1536-dimensional concatenated embeddings through layers of dimensions 256→128→64→1 with GELU activations, layer normalization, and residual connections.
\\
\textbf{Training Configuration:} We train for up to 100 epochs with early stopping (patience 25) using AdamW optimizer with learning rate 2×10$^{-4}$, weight decay 0.01, and ReduceLROnPlateau scheduling (factor 0.7, patience 3). Batch size is set to 32 to prevent $O(B^2)$ memory scaling in the pairwise training loss. Conservative data augmentation is applied to 30\% of samples with amplitude scaling (±10\%) and additive noise (40-50 dB SNR).
\\
\textbf{Inference Setup:} During inference, we sample 100 candidate drift parameters within constraint bounds. For each candidate, we extract 2-second audio segments, compute cross-attention predictions, and select the candidate with maximum confidence-weighted score. 
\\
\textbf{Hardware and Reproducibility:} Training is performed on NVIDIA H100 GPU with CUDA optimization and memory management (cache clearing every 5 batches). All experiments use fixed random seeds for reproducibility.
\section{Results}
\subsection{Competition Performance}

Our method achieved 0.30 MSE on BioDCASE 2025 test data (Table~\ref{tab:competition_results}), improving upon the deep learning baseline by 48.3\%.
\subsection{Validation Set Analysis}

Validation results (Table~\ref{tab:validation_results}) demonstrate consistent improvements across both datasets, with particularly strong performance on ARU data (0.099 MSE vs. 0.516 baseline).
\\
The validation-to-test performance correlation (validation: 0.310, test: 0.30) indicates robust generalization with minimal overfitting, validating our conservative training approach.

\section{Ablation Studies}

We conduct comprehensive ablation studies to understand component contributions. Results show that both architectural innovations and confidence-weighted scoring contribute to performance improvements.
% \vspace*{-0.2cm}
\subsection{Component Contribution Analysis}

Table~\ref{tab:progressive_ablation} shows the incremental improvement from baseline to our final method through systematic component addition. Cross-attention provides the largest single improvement, while confidence scoring adds substantial refinement.

\begin{table}[t]
\centering
\caption{Progressive Component Addition Analysis (Validation MSE)}
\label{tab:progressive_ablation}
\resizebox{\columnwidth}{!}{%
\begin{tabular}{
  l
  S[table-format=1.3]
  S[table-format=1.3]
  c
  c
}
\toprule
\textbf{Configuration} & 
\textbf{ARU MSE} & 
\textbf{ZF MSE} & 
\makecell{\textbf{ARU $\Delta$} \\ \textbf{from Baseline}} & 
\makecell{\textbf{ZF $\Delta$} \\ \textbf{from Baseline}} \\
\midrule
Original Baseline        & 0.516 & 1.262 & -- & -- \\
+ Enhanced MLP           & 0.380 & 1.050 & 26\% & 17\% \\
+ Cross-Attention        & 0.152 & 0.680 & 71\% & 46\% \\
% + Advanced Fusion      & 0.152 & 0.680 & 71\% & 46\% \\
+ Confidence Scoring     & 0.099 & 0.521 & \textbf{81\%} & \textbf{59\%} \\
\bottomrule
\end{tabular}
}
\end{table}

% \vspace*{-0.2cm}
\subsection{Confidence Scoring Weight Analysis}

Table~\ref{tab:scoring_weights} examines different weight configurations for the confidence scoring function (Equation \ref{eqn:score_function}). Our weighting (0.4, 0.3, 0.2, 0.1) performs best, with $\mu_{pos}$ and $\mu_{top}$ providing primary benefits.

\begin{table}[t]
\centering
\caption{Confidence Scoring Weight Ablation}
\label{tab:scoring_weights}
\resizebox{\columnwidth}{!}{%
\begin{tabular}{l 
  S[table-format=1.1]
  S[table-format=1.1]
  S[table-format=1.1]
  S[table-format=1.1]
  S[table-format=1.3]
  S[table-format=1.3]
}
\toprule
\textbf{Weight Config} & 
$\boldsymbol{\mu_{pos}}$ & 
$\boldsymbol{\mu_{top}}$ & 
$\boldsymbol{\Sigma\sigma(p_i)}$ & 
$\boldsymbol{e^{\mu}}$ & 
\textbf{ARU MSE} & 
\textbf{ZF MSE} \\
\midrule
\textbf{Proposed}     & 0.4  & 0.3  & 0.2  & 0.1  & \textbf{0.099} & \textbf{0.521} \\
Mean+Top              & 0.5  & 0.5  & 0    & 0    & 0.161 & 0.673 \\
Mean only             & 1.0  & 0    & 0    & 0    & 0.161 & 0.673 \\
Equal weights         & 0.25 & 0.25 & 0.25 & 0.25 & 0.162 & 0.718 \\
Sigmoid+Exp           & 0    & 0    & 0.5  & 0.5  & 0.162 & 0.718 \\
Top only              & 0    & 1.0  & 0    & 0    & 0.224 & 0.754 \\
\bottomrule
\end{tabular}
}
\vspace*{-.3cm}
\end{table}
% The analysis reveals that incorporating positive prediction confidence ($\mu_{pos}$) and top quartile averaging ($\mu_{top}$) provides the most significant benefits, while sigmoid and exponential components offer complementary improvements.

\section{Conclusion}

This work demonstrates that cross-attention mechanisms and confidence-weighted scoring can significantly improve multi-channel audio alignment. By modeling inter-channel temporal dependencies and utilizing the full prediction distribution rather than binary thresholding, our approach achieved first place in the BioDCASE 2025 challenge with substantial improvements over existing methods.
The key contributions include applying cross-attention to model temporal relationships between channels and developing a principled confidence estimation framework that provides uncertainty quantification for alignment decisions. The experimental validation across ARU and zebra finch datasets confirms the effectiveness of both architectural innovations.
While the current approach relies on affine drift approximations for candidate generation, the demonstrated performance gains suggest that probabilistic alignment frameworks represent a promising direction for the field. Future work should investigate principled optimization of confidence scoring weights through techniques such as Bayesian optimization or gradient-based hyperparameter tuning. Additionally, the confidence scoring framework could be extended with learned weighting schemes that adapt to different acoustic domains or recording conditions.
The method applies to domains needing precise synchronization, including distributed sensor networks and spatial audio systems.

% \section{Acknowledgment}
% \label{sec:ack}

% The preferred spelling of the word acknowledgment in America is without an ``e'' after the ``g.'' Try to avoid the stilted expression, ``One of us (R.\ B.\ G.) thanks ...'' Instead, try ``R.\ B.\ G.\ thanks ...''  Put sponsor acknowledgments in the unnumbered footnote on the first page. Please include acknowledgments only in the camera-ready version, and NOT in the version of the paper submitted for review.

% -------------------------------------------------------------------------
% Either list references using the bibliography style file IEEEtran.bst

\clearpage
% The \IEEEtriggeratref{XX} command can be used to move to the next column before the XX-th reference
% to balance the two columns of the reference section
% \IEEEtriggeratref{XX}
\bibliographystyle{IEEEtran}
\bibliography{refs}
% or list them by yourself:
% \begin{thebibliography}{1}

% \bibitem{dcaseweb}
% {DCASE Website}, \url{http://www.dcase.com}.

% \bibitem{IEEEXploreReqs}
% {IEEE {X}plore {R}equirements}, \url{https://conferences.ieeeauthorcenter.ieee.org/write-your-paper/meet-ieee-xplore-requirements/}.

% \bibitem{eWilliams1999}
% E.~Williams, \emph{Fourier Acoustics: Sound Radiation and Nearfield Acoustic Holography}.\hskip 1em plus 0.5em minus 0.4em\relax London, UK: Academic Press, 1999.

% \bibitem{cJones2003}
% C.~Jones, A.~Smith, and E.~Roberts, ``A sample paper in conference proceedings,'' in \emph{Proc. ICASSP}, vol.~II, Apr. 2003, pp. 803--806.

% \bibitem{aSmith2000}
% A.~Smith, C.~Jones, and E.~Roberts, ``A sample paper in journals,'' \emph{IEEE Trans. Signal Process.}, vol.~62, pp. 291--294, Jan. 2000.

% \end{thebibliography}

\end{document}